\documentclass[11pt]{article}
\usepackage{graphicx}

\def\singleandabitspaced{\baselineskip=\normalbaselineskip\multiply
    \baselineskip by 110\divide\baselineskip by 100}
\def\singlespaced{\baselineskip=\normalbaselineskip}

\newcommand{\centeron}[2]{{\setbox0=\hbox{#1}\setbox1=\hbox{#2}\ifdim
                             \wd1>\wd0\kern.5\wd1\kern-.5\wd0\fi \copy0
                             \kern-.5\wd0\kern-.5\wd1\copy1\ifdim\wd0>\wd1
                             \kern.5\wd0\kern-.5\wd1\fi}}
\newcommand{\ltap}{\>\centeron{\raise.35ex\hbox{$<$}}
                     {\lower.65ex\hbox{$\sim$}}\>}
\newcommand{\gtap}{\>\centeron{\raise.35ex\hbox{$>$}}
                     {\lower.65ex\hbox{$\sim$}}\>}

\def\gev{\rm GeV}

\begin{document}

\singlespaced

\begin{titlepage}

\vspace*{-16mm}
\begin{flushright}
{\small
hep-ph/0411053  \\
}
\end{flushright}

\begin{center}
\vspace*{0.8in}
\mbox{\Large \textbf{Kaluza-Klein Dark Matter, Electrons and}} \\
\vspace*{0.1cm}
\mbox{\Large \textbf{Gamma Ray Telescopes}} \\
\vspace*{1.6cm}
{\large Edward A. Baltz$^1$ and Dan Hooper$^2$} \\
\vspace*{0.5cm}
{\it $^1$ Kavli Institute for Particle Astrophysics and Cosmology, Stanford
University,\\ PO Box 20450, MS 29, Stanford, CA 94309, USA \\
$^2$ Oxford University, Denys Wilkinson Bldg., Keble Road, Oxford OX1-3RH, UK}
\\
\vspace*{0.6cm}
{\tt eabaltz@slac.stanford.edu, hooper@astro.ox.ac.uk} \\
\vspace*{1.5cm}
\end{center}

\begin{abstract} 

Kaluza-Klein dark matter particles can annihilate efficiently into
electron-positron pairs, providing a discrete feature (a sharp edge) in the
cosmic $e^+ e^-$ spectrum at an energy equal to the particle's mass (typically
several hundred GeV to one TeV).  Although this feature is probably beyond the
reach of satellite or balloon-based cosmic ray experiments (those that
distinguish the charge and mass of the primary particle), gamma ray telescopes
may provide an alternative detection method.  Designed to observe very
high-energy gamma-rays, ACTs also observe the diffuse flux of electron-induced
electromagnetic showers.  The GLAST satellite, designed for gamma ray
astronomy, will also observe any high energy showers (several hundred GeV and
above) in its calorimeter.  We show that high-significance detections of an
electron-positron feature from Kaluza-Klein dark matter annihilations are
possible with GLAST, and also with ACTs such as HESS, VERITAS or MAGIC.

\end{abstract}

\end{titlepage}

\newpage
\setcounter{page}{2}

\singleandabitspaced

\section{Introduction}

The identity of dark matter remains one of the primary outstanding puzzles of
modern astrophysics~\cite{review}. The numerous planned and ongoing searches
for particle dark matter include collider experiments \cite{collider} as well
as direct \cite{direct} and indirect detection efforts.

Indirect dark matter searches attempt to identify the products of dark matter
annihilations produced in regions such as the galactic halo, the galactic
center or the center of the Sun. Such annihilation products include gamma-rays
\cite{gammarays}, neutrinos \cite{neutrinos}, anti-protons
\cite{antiprotons,antimatter}, anti-deuterons \cite{antimatter,antideuterons}
and positrons \cite{antimatter,hoopersilk,positrons,posbaltz,kkpos}.

One interesting dark matter candidate are Kaluza-Klein (KK) states present in
models with extra spatial dimensions. In particular, in models of Universal
Extra Dimensions (UED) \cite{universal}, in which all of the Standard Model
fields are allowed to propagate in the bulk, it has been shown that the
Lightest Kaluza-Klein Particle (LKP) can be stable and a viable dark matter
candidate \cite{taitservant,kkdark,kkdarkfeng}. In such a scenario, the LKP may
be stabilized as a result of momentum conservation in the extra dimensions (KK
number conservation). For a UED model to be phenomenologically viable, however,
the extra dimensions must be modded out by an orbifold, which can lead to the
violation of KK number conservation. A symmetry, called KK-parity, may remain,
however, which insures that the LKP cannot decay, in much the same way that
R-parity stabilizes the lightest supersymmetric state in many models. A natural
choice for the LKP is the first KK excitation of the hypercharge gauge
boson. We will refer to this state simply as the LKP, or as Kaluza-Klein Dark
Matter (KKDM).

In this article, we focus on detecting electrons and positrons which are
produced in KKDM annihilations in the galactic halo.  Such measurements are
particularly useful for KKDM searches, as their annihilations often produce
$e^+ e^-$ pairs directly, resulting in a dramatic feature in the spectrum at an
energy equal to the WIMP mass~\cite{kkpos,kkdarkfeng}.

Measurements of the cosmic $e^+ e^-$ spectrum have been made by the HEAT
experiment \cite{heat} and will be studied with much greater precision in the
future with PAMELA and AMS-02 \cite{pamelaams}.  None of these experiments can
accurately measure this spectrum above $\sim$200 GeV, however.  Higher energy
particles will be detected, but not easily identified and characterized.  This
is unfortunate, as KKDM is constrained by electroweak precision measurements to
be heavier than about 300 GeV \cite{ew300}. To look for the an injection of
electrons and positrons at energies above 300 GeV, other techniques need to be
pursued. In this article, we suggest using high energy gamma ray telescopes,
both the planned GLAST satellite and ground based Atmospheric Cerenkov
Telescopes (ACTs), to search for signatures of Kaluza-Klein Dark Matter in the
cosmic electron-positron spectrum.

\section{Electrons and Positrons From Kaluza-Klein Dark Matter}

Kaluza-Klein Dark Matter (KKDM) annihilates through very different modes than
many other WIMP candidates, such as neutralinos in supersymmetric
models. Neutralino annihilations to fermions are chirality suppressed by a
factor of $m^2_{\rm{f}}/m^2_{\chi}$, and thus produce essentially no $e^+ e^-$
pairs directly. KKDM, being a boson, is not similarly suppressed and can
annihilate directly to $e^+e^-$, $\mu^+ \mu^-$ and $\tau^+ \tau^-$, each of
which yield a generous number of high energy electrons and positron. The KKDM
annihilation cross section is proportional to the hypercharge of the final
state fermions to the fourth power, and thus most annihilations produce pairs
of charged leptons (approximately $20\%$ per generation). Other dominant modes
include annihilations to up-type quarks (approximately $11\%$ per generation),
neutrinos (approximately $1.2\%$ per generation), Higgs bosons (approximately
$2.3\%$) and down-type quarks (approximately $0.7\%$ per generation). The total
annihilation cross section of KKDM is given by
\begin{equation}
\langle\sigma v\rangle = \frac{95 g_1^4}{324 \pi m^2_{\rm LKP}} \simeq
\frac{1.7 \times 10^{-26} \, \rm{cm^3}/\rm{s}}{\left(m_{\rm
LKP}/\rm{TeV}\right)^2}.
\label{cross}
\end{equation}

If no other Kaluza-Klein states play a significant role in the thermal
freeze-out of the LKP, Eq.~\ref{cross} requires a LKP with a mass of about
700-1000 GeV in order to produce the quantity of cold dark matter measured by
WMAP \cite{wmap}. It has been shown, however, that if other Kaluza-Klein states
are only slightly more heavy than the LKP, they may freeze-out
quasi-independently, eventually decaying into LKPs and enhancing the KKDM relic
density non-thermally \cite{taitservant}. In this case, lighter LKPs can make
up all of the measured cold dark matter density.

We have used PYTHIA \cite{pythia}, as implemented in the DarkSUSY program
\cite{darksusy}, to calculate the $e^+ e^-$ spectrum generated in KKDM
annihilations. The spectrum injected is quite different from that observed at
Earth, however. Electrons and positrons travel through the galactic environment
under the influence of tangled interstellar magnetic fields and lose energy via
inverse Compton and synchrotron interactions. These effects can be modeled by
the diffusion-loss equation:
\begin{eqnarray}
\frac{\partial}{\partial t}\frac{dn_{e^{\pm}}}{dE_{e^{\pm}}} =
\vec{\nabla} \cdot \bigg[K(E_{e^{\pm}},\vec{x})
\vec{\nabla} \frac{dn_{e^{\pm}}}{dE_{e^{\pm}}} \bigg] +
\frac{\partial}{\partial E_{e^{\pm}}}
\bigg[b(E_{e^{\pm}},\vec{x})\frac{dn_{e^{\pm}}}{dE_{e^{\pm}}} \bigg] +
Q(E_{e^{\pm}},\vec{x}),
\label{dif}
\end{eqnarray}
where $dn_{e^{\pm}}/dE_{e^{\pm}}$ is the number density of electrons and
positrons per unit energy, $K(E_{e^{\pm}},\vec{x})$ is the diffusion constant,
$b(E_{e^{\pm}},\vec{x})$ is the rate of energy loss and
$Q(E_{e^{\pm}},\vec{x})$ is the source term.
 
We use a diffusion constant of $K(E_{e^{\pm}}) = 3.3 \times 10^{28}
\bigg[3^{0.47} + (E_{e^{\pm}}/\gev)^{0.47} \bigg] \,\rm{cm}^2 \,
\rm{s}^{-1}$~\cite{diffusion}, and an energy loss rate of $b(E_{e^{\pm}}) =
10^{-16} (E_{e^{\pm}}/\gev)^2 \,\, \gev\,\,\rm{s}^{-1}$, which is the result of
inverse Compton scattering on starlight and the cosmic microwave background,
and synchrotron radiation due to the galactic magnetic field \cite{lossrate}.
We assume that the diffusion zone is a slab of thickness $2L$, taking
$L=4\,$kpc, and we apply free escape boundary conditions.

The source term, $Q(E_{e^{\pm}})$, is determined by the electron-positron
spectrum injected per annihilation and by the annihilation
rate, which normalizes the flux. The annihilation rate depends on both the KKDM
annihilation cross section (Eq.~\ref{cross}) and the galactic distribution of
dark matter. For our halo dark matter distribution, we use an NFW profile
\cite{nfw}, although our results are not highly dependent on this choice
\cite{hoopersilk,posbaltz}. In addition to this choice, the degree of
inhomogeneity or substructure in the dark matter distribution can effect the
annihilation rate. This is often parameterized by a quantity called the {\it
boost factor}. This quantity is essentially the average of the square of the
dark matter density over the square of the average of the dark matter
density. Equivalently, it is the factor the annihilation rate is enhanced as a
result of dark matter clumping. Typical values of the boost factor are on the
scale of 2 to 5. Values much larger than this require very large amounts of
dark substructure and are highly unnatural \cite{hooperpos}.

While solutions to the propagation equation are complex, it is fairly simple to
understand the feature we are most interested in, namely the magnitude of the
edge feature in the spectrum.  We take $Q_{\rm line}(m_{\rm
KKDM},\vec{x}_\odot)$ (in cm$^{-3}$ s$^{-1}$) as the rate of electron and
positron injection from direct annihilation to $e^+e^-$ locally.  The spectrum
near the edge is then simply
\begin{equation}
\frac{dn_{e^{\pm}}}{dE_{e^{\pm}}}=\frac{Q_{\rm line}(m_{\rm
KKDM},\vec{x}_\odot)}{b(E_{e^{\pm}},\vec{x}_\odot)}\,\theta\left(m_{\rm
KKDM}-E_{e^\pm}\right).
\end{equation}

\begin{figure}[thb]
\vbox{\kern2.4in\includegraphics{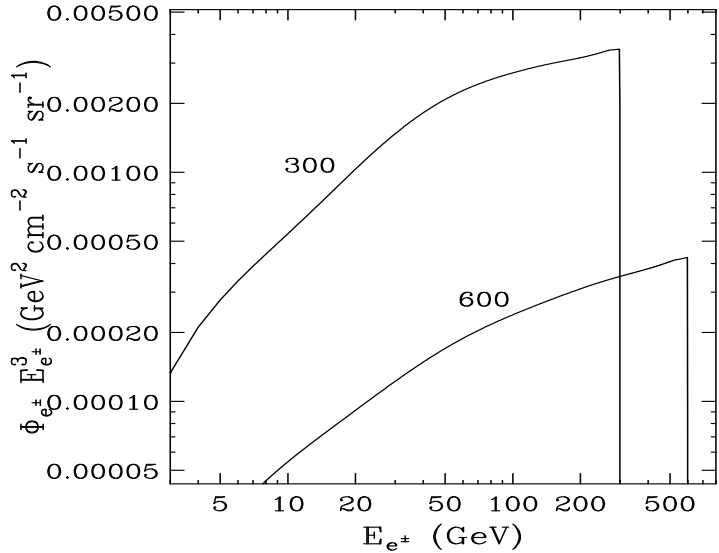}}
\caption{The spectrum of electrons plus positrons, including the effects of
propagation, from Kaluza-Klein Dark Matter (KKDM) annihilations. Annihilations
of KKDM produce equal fractions of $\tau^+ \tau^-$, $\mu^+ \mu^-$ and $e^+ e^-$
pairs (approximately $20\%$ each) as well as up-type quarks (approximately
$11\%$ per generation), neutrinos (approximately $1.2\%$ per generation), Higgs
bosons (approximately $2.3\%$) and down-type quarks (approximately $0.7\%$ per
generation). Results for KKDM with masses of 300 and 600 GeV are shown. An NFW
dark matter distribution with a boost factor of 5 and $\rho_{\rm{local}}=0.4
\,\rm{GeV/cm^3}$ was used.}
\label{fluxKK}
\end{figure}

We show the spectrum of electrons plus positrons from Kaluza-Klein dark matter
annihilations after propagation in figure~\ref{fluxKK}.  In the next section,
we discuss the prospects for gamma ray telescopes, both ACTs and GLAST, to
detect this flux.

\vspace{1.0cm}

\vspace{1.0cm}

\section{Prospects for Gamma Ray Telescopes}

Atmospheric Cerenkov Telescopes (ACTs) are ground based experiments designed to
detect very high energy gamma-rays by imaging the Cerenkov light produced in
air showers generated in the atmosphere.  Currently, ACTs have made gamma-ray
measurements in the energy range of roughly 200 GeV to 10 TeV, although
thresholds as low as $\sim$50 GeV may be possible with future technology.  The
weakness of ACTs is in their limited ability to identify the primary particle
that induced the shower.  While it is usually possible (typically with better
than 99\% confidence) to distinguish hadronic showers (from e.g.\ primary
protons or heavier nuclei) from electromagnetic showers, further identification
of the primary cosmic ray cannot be achieved.  In particular, showers caused by
primary gamma-rays, electrons, and positrons are indistinguishable to ACTs.
These instruments are only useful for gamma-ray astronomy because of the lack
of point sources of cosmic electrons.

Modern ACTs, with effective areas on the order of $10^5$ square meters and
fields-of-view of a few degrees, may provide a useful window into the diffuse
cosmic ray spectrum in addition to their role as gamma-ray telescopes. In this
section, we assess the prospects for ACTs, such as HESS \cite{hess}, VERITAS
\cite{veritas} and MAGIC \cite{magic}, to observe a sudden drop in the
electron-positron spectrum (see figure~\ref{fluxKK}) which would be predicted
for Kaluza-Klein dark matter.

The GLAST satellite is designed for gamma ray astronomy in the energy range
between about 20 MeV and 300 GeV \cite{GLAST}.  Higher energy photons will be
measured, but the energy resolution degrades significantly (perhaps as bad as
50\%).  In addition, any high energy (above a few hundred GeV) showers in the
calorimeter will be recorded, regardless of primary (electromagnetic or
hadronic).  This capability may allow a detection of the electron-positron
edge.

It is expected that the flux of charged cosmic rays is (at least roughly)
isotropic, thus we are searching for an all-sky signature.  We propose to
consider the entire datasets of ACTs or GLAST, without regard for position on
the sky except where necessary for e.g.\ energy calibration as a function of
zenith angle.  Point sources (presumably of gamma-rays) may be excised from the
dataset without significant effect, as the angular resolution of these
instruments is typically a fraction of a degree while the field-of-view is
several degrees.  All other showers will be associated with the cosmic
backgrounds of hadrons, electrons, positrons and gamma-rays.  Considering the
summed energy spectrum of all of these backgrounds, a sharp edge in the
spectrum of electrons and positrons would become evident with enough exposure.

To assess the sensitivity of these gamma ray telescopes to a flux of electrons
and positrons from Kaluza-Klein dark matter annihilations, we first must
estimate the relevant background rates.

\begin{itemize}
\item
The cosmic ray electron spectrum over the GeV-TeV energy range is given by:
\begin{equation}
\frac{dN_e}{dE_e} \simeq 0.07 \times \, \bigg( \frac{E_e}{1\,\rm{GeV}} \bigg)^{-3.3}  \rm{GeV}^{-1}\,\rm{cm}^{-2}\,\rm{sec}^{-1} \, \rm{sr}^{-1}.
\end{equation}
\item
The hadronic cosmic ray spectrum over the same range is:
\begin{equation}
\frac{dN_{\rm{had}}}{dE_{\rm{had}}} \simeq 3 \times \, \bigg(
\frac{E_e}{1\,\rm{GeV}} \bigg)^{-2.7}
\rm{GeV}^{-1}\,\rm{cm}^{-2}\,\rm{sec}^{-1} \, \rm{sr}^{-1}.
\end{equation}
\item
The diffuse gamma-ray background has the same spectral shape but is
considerably smaller:
\begin{equation}
\frac{dN_{\gamma}}{dE_{\gamma}} \simeq 4 \times 10^{-4} \times \, \bigg(
\frac{E_e}{1\,\rm{GeV}} \bigg)^{-2.7}
\rm{GeV}^{-1}\,\rm{cm}^{-2}\,\rm{sec}^{-1} \, \rm{sr}^{-1}.
\end{equation}
\end{itemize}

Although showers (whether atmospheric, or in the GLAST calorimeter) produced by
gamma-rays cannot be distinguished from electron induced showers (or vice
versa), the vast majority of hadronic showers can be identified and
rejected. The total background from these components is then:
\begin{equation}
\frac{dN_{\rm{bg}}}{dE_{\rm{bg}}} \simeq (3 \times \epsilon_{\rm{had}} + 4
\times 10^{-4}) \times \, \bigg( \frac{E_e}{1\,\rm{GeV}} \bigg)^{-2.7} + 0.07
\times \, \bigg( \frac{E_e}{1\,\rm{GeV}} \bigg)^{-3.3} \,
\rm{GeV}^{-1}\,\rm{cm}^{-2}\,\rm{sec}^{-1} \, \rm{sr}^{-1},
\label{bg}
\end{equation}
where $\epsilon_{\rm{had}}$ is the fraction of hadronic showers which cannot be
rejected. We can see that the diffuse gamma-ray component of the background is
important only if the experiment's hadron rejection efficiency is greater than
about $99.99\%$.  For $\epsilon_{\rm{had}}=0.01$ (appropriate to ACTs), the
electron background is irrelevant.  For better rejection
($\epsilon_{\rm{had}}=0.001-0.0001$, appropriate to GLAST), the hadron / photon
background dominates above a few hundred GeV.

In order for a gamma ray telescope to identify the presence of an
electron-positron feature from Kaluza-Klein dark matter annihilations, there
must be a statistically significant variation in the spectrum at the energy,
$E=m_{\rm{KKDM}}$. Considering an energy bin of width corresponding to the
energy resolution of the experiment, the statistical significance of the
feature is $\simeq \frac{S}{\sqrt{N}}$, where $S$ and $N$ are the numbers of
signal and background events in the bin. $N$ is determined by integrating
Eq.~\ref{bg} over the energy bin width, while $S$ is the integrated result of
Eq.~\ref{dif} along with the annihilation rate and injected spectrum (see
Fig.~\ref{fluxKK}).

\begin{figure}[thb]
\vbox{\kern2.8in\includegraphics{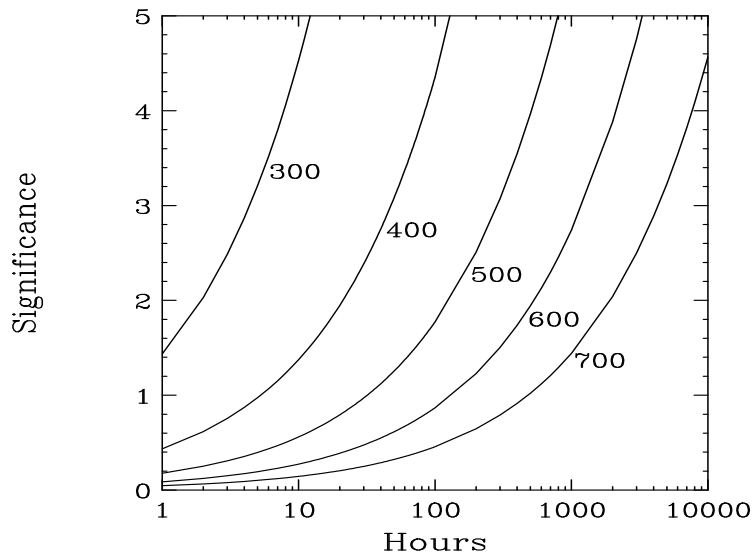}}
\caption{The significance of the $e^{\pm}$ feature from Kaluza-Klein dark
matter annihilations at $E=m_{\rm{KKDM}}$ in a modern Atmospheric Cerenkov
Telescope (ACT), such as HESS, VERITAS or MAGIC, as a function of time
observed. We have considered an ACT with $15\%$ energy resolution, $99\%$
hadronic rejection, a 0.003 sr field-of-view and $2\times 10^5$ square meters
effective area. Results for dark matter masses of 300, 400, 500, 600 and 700
GeV are shown. An NFW dark matter distribution with a boost factor of 5 and
$\rho_{\rm{local}}=0.4 \,\rm{GeV/cm^3}$ was used.  To compare with GLAST, the
exposure time axis should be multiplied by a factor of 16, thus the 600 GeV
particle detected at 5$\sigma$ in 3000 hours with an ACT requires 48,000 hours
with GLAST.  Both datasets could be collected in roughly 7 years.}
\label{sig}
\end{figure}

Our estimate of the sensitivity of an ACT to Kaluza-Klein dark matter
annihilations is shown in figure~\ref{sig}. We have considered an ACT with
$15\%$ energy resolution (thus a energy bin of width $\Delta E \approx 0.3\,
m_{\rm{KKDM}}$), $99\%$ hadronic rejection ($\epsilon_{\rm had}=0.01$), a 0.003
sr field-of-view and $2 \times 10^5$ square meters of effective area.  These
are reasonable values for state-of-the-art ACTs such as HESS, VERITAS and
MAGIC. We normalize the KKDM annihilation rate using a boost factor of 5, a
local dark matter density of $\rho_{\rm{local}}=0.4 \,\rm{GeV/cm^3}$ and the
annihilation cross section of Eq.~\ref{cross}.

We see from figure~\ref{sig} that after only about 10 hours of observation, a
300 GeV KKDM particle could be detected with 5$\sigma$ significance. A hundred
hours would be needed to reach similar sensitivity for KKDM with a mass of
about 400 GeV. To reach a 600 GeV KKDM particle, 3000 hours would be required.
Due to requirements of good weather and dark (moonless or nearly so) nights,
the duty cycle of an ACT is typically only 5-7\%, or 450--600 hours of
observation per year.  We thus consider several thousand total hours of
observation a plausible goal, especially if the exposure of multiple
experiments can be combined.

In comparison, the capabilities of GLAST seem quite different.  While we have
assumed ACTs with an exposure of 600 m$^2$ sr and a 5\% duty cycle, the GLAST
calorimeter will have an exposure of 7.5 m$^2$ sr and an 80\% duty cycle.
Thus, in a given time period, the ACTs have roughly 5 times the exposure.
GLAST has an advantage, in that the rejection of hadronic showers should be
better than 99.9\%, and perhaps as good as 99.99\%, which is the useful limit
as the gamma ray background comes in at this level.  The signal to noise to
detect the edge in the electron plus positron spectrum should improve as ${\rm
S/N}\propto1/\sqrt{\epsilon_{\rm had}}$, namely GLAST should gain a factor of
between 3 and 10 in sensitivity due to the improved hadron rejection.  This
factor roughly makes up for the fact that GLAST will have a factor of 5 less
exposure.  Lastly, we can assume an integration over the same energy bin, 30\%
wide.

Finally, we will compare the sensitivity of these experiments to those designed to observe cosmic positrons, such as PAMELA or AMS-02. PAMELA and AMS-02 will accumulate considerably smaller exposures than GLAST, $20.5$ and $450$ cm$^2$ sr, respectively. Even if they achieve their projected hadronic rejection on the order of 99.9\% \cite{amsreject}, they will not be capable of competing with GLAST or ACTs in the energy range we consider here.

%\vspace{1.0cm}

%\vspace{0.0cm}

\clearpage

\section{Conclusions}

Electrons and positrons produced directly in Kaluza-Klein Dark Matter (KKDM)
annihilations can result in a discontinuity in the diffuse spectrum observed by
gamma ray telescopes both on the ground (ACTs such as HESS, VERITAS, or MAGIC)
and in space (GLAST).  We have shown that this feature can be observed at
statistically significant levels in either ACTs or GLAST for KKDM particles
with masses of up to 600 GeV, if several years are spent accumulating data.

\vspace{0.5cm}

\section*{Acknowledgements}

We would like to thank Bill Atwood, Daniel Ferenc, Peter Michelson, Ignacio de
la Calle Perez, Roger Romani, and David A. Smith for numerous helpful
discussions.  We would also like to thank the organizers of the IDM 2004
conference in Edinburgh, where this research was begun.  EAB was supported by
the U.S. Department of Energy under contract number \mbox{DE-AC02-76SF00515}.
DH is supported by the Leverhulme trust.


\begin{thebibliography}{99}
\singlespaced

%%%%



\bibitem{review}
For a recent review, see G.~Bertone, D.~Hooper and J.~Silk,
%``Particle dark matter: Evidence, candidates and constraints,''
Physics Reports, in press, arXiv:hep-ph/0404175.
%%CITATION = HEP-PH 0404175;%%


\bibitem{collider}
R.~Barate {\it et al.}  [ALEPH Collaboration],
  %``Search for supersymmetric particles in e+ e- collisions at s**(1/2
  Phys.\ Lett.\ B {\bf 499}, 67 (2001);
  %%[arXiv:hep-ex/0011047];
  %%CITATION = HEP-EX 0011047;%%
 G.~Abbiendi {\it et al.}  [OPAL Collaboration],
  %``Search for anomalous production of acoplanar di-lepton events in 
  Eur.\ Phys.\ J.\ C {\bf 14}, 51 (2000);
  %%[arXiv:hep-ex/9909052].
  %%CITATION = HEP-EX 9909052;%%
 M.~Acciarri {\it et al.}  [L3 Collaboration],
  %``Search for scalar leptons in e+ e- collisions at s**(1/2) = 189-GeV,''
  Phys.\ Lett.\ B {\bf 471}, 280 (1999);
  %%[arXiv:hep-ex/9910006].
  %%CITATION = HEP-EX 9910006;%%
 A.~Heister {\it et al.}  [ALEPH Collaboration],
  %``Search for scalar leptons in e+ e- collisions at centre-of-mass
  Phys.\ Lett.\ B {\bf 526}, 206 (2002);
  %%[arXiv:hep-ex/0112011].
  %%CITATION = HEP-EX 0112011;%%
 D.~Acosta {\it et al.}  [CDF Collaboration],
  %``Search for single top quark production in p anti-p collisions at
  Phys.\ Rev.\ D {\bf 65}, 091102 (2002);
  %%[arXiv:hep-ex/0110067].
  %%CITATION = HEP-EX 0110067;%%
 F.~Abe {\it et al.}  [CDF Collaboration],
  %``Search for gluinos and squarks at the Fermilab Tevatron collider,''
  Phys.\ Rev.\ D {\bf 56}, 1357 (1997);
 B.~Abbott {\it et al.}  [D0 Collaboration],
  %``Search for bottom squarks in p anti-p collisions at s**(1/2) = 1.8-TeV,''
  Phys.\ Rev.\ D {\bf 60}, 031101 (1999);
  %%[arXiv:hep-ex/9903041].
  %%CITATION = HEP-EX 9903041;%%
 B.~Abbott {\it et al.}  [D0 Collaboration],
  %``A search for dilepton signatures from minimal low-energy supergravit
  Phys.\ Rev.\ D {\bf 63}, 091102 (2001);
  %%CITATION = PHRVA,D63,091102;%%
 S.~Abel {\it et al.}  [SUGRA Working Group Collaboration],
  %``Report of the SUGRA working group for run II of the Tevatron,''
  hep-ph/0003154;
  %%CITATION = HEP-PH 0003154;%%
A.~Birkedal, K.~Matchev and M.~Perelstein,
%``Dark matter at colliders: A model-independent approach,''
arXiv:hep-ph/0403004.
%%CITATION = HEP-PH 0403004;%%

\bibitem{direct}
A.~Drukier and L.~Stodolsky,
%``Principles And Applications Of A Neutral Current Detector For Neutrino Physics And Astronomy,''
Phys.\ Rev.\ D {\bf 30}, 2295 (1984);
%%CITATION = PHRVA,D30,2295;%%
M.~W.~Goodman and E.~Witten,
%``Detectability Of Certain Dark-Matter Candidates,''
Phys.\ Rev.\ D {\bf 31}, 3059 (1985);
%%CITATION = PHRVA,D31,3059;%%
H.~Baer and M.~Brhlik,
%``Neutralino dark matter in minimal supergravity: Direct detection vs. 
Phys.\ Rev.\ D {\bf 57}, 567 (1998);
%%CITATION = HEP-PH 9706509;%%
G.~Servant and T.~M.~P.~Tait,
%``Elastic scattering and direct detection of Kaluza-Klein dark matter,''
New J.\ Phys.\  {\bf 4}, 99 (2002)
[arXiv:hep-ph/0209262].
%%CITATION = HEP-PH 0209262;%%



\bibitem{gammarays}
F~ W.~Stecker and A.~J.~Tylka, 
Astrophys.\ J. 343, 169 (1989);
F.~W.~Stecker, Phys.\ Lett.\ B 201, 529 (1988);
L.~Bergstrom, J.~Edsjo and P.~Ullio,
%``Spectral gamma-ray signatures of cosmological dark matter  annihilations,''
Phys.\ Rev.\ Lett.\  {\bf 87}, 251301 (2001);
H.~U.~Bengtsson, P.~Salati and J.~Silk,
%``Quark Flavors And The Gamma-Ray Spectrum From Halo Dark Matter 
Nucl.\ Phys.\ B {\bf 346}, 129 (1990);
%%CITATION = NUPHA,B346,129;%%
V.~Berezinsky, A.~Bottino and G.~Mignola,
%``High-energy gamma radiation from the galactic center due to
Phys.\ Lett.\ B {\bf 325}, 136 (1994)
[arXiv:hep-ph/9402215];
%%CITATION = HEP-PH 9402215;%%
D.~Hooper and B.~L.~Dingus,
%``Limits on supersymmetric dark matter from EGRET observations of the
arXiv:astro-ph/0210617;
%%CITATION = ASTRO-PH 0210617;%%
L.~Bergstrom, P.~Ullio and J.~H.~Buckley,
%``Observability of gamma rays from dark matter neutralino annihilations 
Astropart.\ Phys.\  {\bf 9}, 137 (1998);
%%CITATION = ASTRO-PH 9712318;%%
G.~Bertone, G.~Servant and G.~Sigl,
%``Indirect detection of Kaluza-Klein dark matter,''
Phys.\ Rev.\ D {\bf 68}, 044008 (2003)
[arXiv:hep-ph/0211342];
%%CITATION = HEP-PH 0211342;%%
L.~Bergstrom, T.~Bringmann, M.~Eriksson and M.~Gustafsson,
arXiv:astro-ph/0410359.


\bibitem{neutrinos}
J.~Silk, K.~Olive and M.~Srednicki,
%``The photino, the sun, and high-energy neutrinos,''
Phys.\ Rev.\ Lett.\ {\bf 55}, 257 (1985);
J.~S.~Hagelin, K.~W.~Ng, K.~A.~Olive, 
Phys.\ Lett.\ B {\bf 180}, 375 (1986); 
K.~Freese,
%``Can Scalar Neutrinos Or Massive Dirac Neutrinos Be The Missing Mass?,''
Phys.\ Lett.\ B {\bf 167}, 295 (1986);
L.~M.~Krauss, M.~Srednicki and F.~Wilczek,
%``Solar System Constraints And Signatures For Dark Matter Candidates,''
Phys.\ Rev.\ D {\bf 33}, 2079 (1986);
T.~K.~Gaisser, G.~Steigman and S.~Tilav,
%``Limits On Cold Dark Matter Candidates From Deep Underground Detectors,''
Phys.\ Rev.\ D {\bf 34}, 2206 (1986);
V.~D.~Barger, F.~Halzen, D.~Hooper and C.~Kao,
%``Indirect search for neutralino dark matter with high energy neutrinos,''
Phys.\ Rev.\ D {\bf 65}, 075022 (2002);
%%CITATION = HEP-PH 0105182;%%
D.~Hooper and G.~D.~Kribs,
%``Probing Kaluza-Klein dark matter with neutrino telescopes,''
Phys.\ Rev.\ D {\bf 67}, 055003 (2003)
[arXiv:hep-ph/0208261];
%%CITATION = HEP-PH 0208261;%%
L.~Bergstrom, J.~Edsjo and P.~Gondolo,
%``Indirect detection of dark matter in km-size neutrino telescopes,''
Phys.\ Rev.\ D {\bf 58}, 103519 (1998).


\bibitem{antiprotons}
F.~W.~Stecker, S.~Rudaz and T.~F.~Walsh,  
Phys.\ Rev.\ Letters {\bf 55}, 2622 (1985);
S.~Rudaz and F.~W.~Stecker, 
Astrophys.\ J. 325, 16 (1988);
L.~Bergstrom, J.~Edsjo and P.~Ullio,
%``Cosmic antiprotons as a probe for neutralino dark matter?,''
arXiv:astro-ph/9906034;
%%CITATION = ASTRO-PH 9906034;%%
A.~Bottino, F.~Donato, N.~Fornengo and P.~Salati,
%``Which fraction of the measured cosmic ray antiprotons might be due to 
Phys.\ Rev.\ D {\bf 58}, 123503 (1998);
%%CITATION = ASTRO-PH 9804137;%%
F.~Donato, N.~Fornengo, D.~Maurin, P.~Salati and R.~Taillet,
%``Antiprotons in cosmic rays from neutralino annihilation,''
arXiv:astro-ph/0306207.
%%CITATION = ASTRO-PH 0306207;%%

\bibitem{antimatter}
S.~Profumo and P.~Ullio,
%``The role of antimatter searches in the hunt for supersymmetric 
JCAP {\bf 0407}, 006 (2004)
[arXiv:hep-ph/0406018].
%%CITATION = HEP-PH 0406018;%%

\bibitem{antideuterons}
F.~Donato, N.~Fornengo and P.~Salati,
%``Antideuterons as a signature of supersymmetric dark matter,''
Phys.\ Rev.\ D {\bf 62}, 043003 (2000)
[arXiv:hep-ph/9904481].
%%CITATION = HEP-PH 9904481;%%

\bibitem{hoopersilk}
D.~Hooper and J.~Silk,
%``Searching for dark matter with future cosmic positron experiments,''
arXiv:hep-ph/0409104.
%%CITATION = HEP-PH 0409104;%%


\bibitem{positrons}
M.~Kamionkowski and M.~S.~Turner,
%``A Distinctive Positron Feature From Heavy Wimp Annihilations In The
Phys.\ Rev.\ D {\bf 43}, 1774 (1991);
%%CITATION = PHRVA,D43,1774;%%
E.~A.~Baltz, J.~Edsjo, K.~Freese and P.~Gondolo,
%``The positron excess and supersymmetric dark matter,''
arXiv:astro-ph/0211239;
%%CITATION = ASTRO-PH 0211239;%%
M.~S.~Turner and F.~Wilczek,
Phys.\ Rev.\ D, {\bf 42}, 1001 (1990);
A.~J.~Tylka,
Phys.\ Rev.\ Lett., {\bf 63}, 840 (1989);
G.~L.~Kane, L.~T.~Wang and T.~T.~Wang,
Phys.\ Lett.\ B {\bf 536}, 263 (2002);
J.~L.~Feng, K.~T.~Matchev and F.~Wilczek,
%``Prospects for indirect detection of neutralino dark matter,''
Phys.\ Rev.\ D {\bf 63}, 045024 (2001);
E.~A.~Baltz, J.~Edsjo, K.~Freese and P.~Gondolo,
%``The cosmic ray positron excess and neutralino dark matter,''
Phys.\ Rev.\ D {\bf 65}, 063511 (2002);
G.~L.~Kane, L.~T.~Wang and J.~D.~Wells,
%``Supersymmetry and the positron excess in cosmic rays,''
Phys.\ Rev.\ D {\bf 65}, 057701 (2002);
%%CITATION = HEP-PH 0108138;%%

\bibitem{posbaltz}
E.~A.~Baltz and J.~Edsjo,
%``Positron propagation and fluxes from neutralino annihilation in the  halo,''
Phys.\ Rev.\ D {\bf 59} (1999) 023511
[arXiv:astro-ph/9808243].
%%CITATION = ASTRO-PH 9808243;%%

\bibitem{kkpos}
D.~Hooper and G.~D.~Kribs,
%``Kaluza-Klein dark matter and the positron excess,''
arXiv:hep-ph/0406026.
%%CITATION = HEP-PH 0406026;%%


\bibitem{universal}
I.~Antoniadis,
%``A Possible New Dimension At A Few Tev,''
Phys.\ Lett.\ B {\bf 246}, 377 (1990);
%%CITATION = PHLTA,B246,377;%%
I.~Antoniadis, K.~Benakli and M.~Quir\'os,
%``Production of Kaluza-Klein states at future colliders,''
Phys.\ Lett.\ B {\bf 331}, 313 (1994)
[arXiv:hep-ph/9403290];
%%CITATION = HEP-PH 9403290;%%
H.~C.~Cheng, K.~T.~Matchev and M.~Schmaltz,
%``Bosonic supersymmetry? Getting fooled at the LHC,''
Phys.\ Rev.\ D {\bf 66}, 056006 (2002)
[arXiv:hep-ph/0205314].
%%CITATION = HEP-PH 0205314;%%

\bibitem{taitservant}
G.~Servant and T.~M.~P.~Tait,
%``Is the lightest Kaluza-Klein particle a viable dark matter candidate?,''
Nucl.\ Phys.\ B {\bf 650}, 391 (2003)
[arXiv:hep-ph/0206071].
%%CITATION = HEP-PH 0206071;%%


\bibitem{kkdark}
E.~W.~Kolb and R.~Slansky,
 %``Dimensional Reduction In The Early Universe: Where Have The Massive
%Particles Gone?,''
Phys.\ Lett.\ B {\bf 135}, 378 (1984);
%%CITATION = PHLTA,B135,378;%%
K.~R.~Dienes, E.~Dudas and T.~Gherghetta,
%``Grand unification at intermediate mass scales through extra dimensions,''
Nucl.\ Phys.\ B {\bf 537}, 47 (1999)
[arXiv:hep-ph/9806292].
%%CITATION = HEP-PH 9806292;%%

\bibitem{kkdarkfeng}
H.~C.~Cheng, J.~L.~Feng and K.~T.~Matchev,
%``Kaluza-Klein dark matter,''
Phys.\ Rev.\ Lett.\  {\bf 89}, 211301 (2002)
[arXiv:hep-ph/0207125].
%%CITATION = HEP-PH 0207125;%%

\bibitem{heat}
S.~W.~Barwick {\it et al.}  [HEAT Collaboration],
%``Measurements of the cosmic-ray positron fraction from 1-GeV to 50-GeV,''
Astrophys.\ J.\  {\bf 482}, L191 (1997)
[arXiv:astro-ph/9703192];
%%CITATION = ASTRO-PH 9703192;%%
S.~Coutu {\it et al.}  [HEAT-pbar Collaboration],
``Positron Measurements With the HEAT-pbar Instrument'',
in Proceedings of 27th ICRC (2001);
S.~Coutu {\it et al.},
Astropart.\ Phys.\ {\bf 11}, 429 (1999), 
[arXiv:astro-ph/9902162].



\bibitem{pamelaams}
M.~Circella  [PAMELA Collaboration],
%``The Space Mission Pamela,''
Nucl.\ Instrum.\ Meth.\ A {\bf 518}, 153 (2004);
%%CITATION = NUIMA,A518,153;%%
P.~Spillantini,
%``The Balloon-Borne And Pamela Experiments For The Study Of The Antimatter Component In Cosmic Rays,''
Nucl.\ Instrum.\ Meth.\ B {\bf 214}, 116 (2004);
%%CITATION = NUIMA,B214,116;%%
P.~Picozza and A.~Morselli,
%``Antimatter research in space,''
J.\ Phys.\ G {\bf 29}, 903 (2003)
[arXiv:astro-ph/0211286].
%%CITATION = ASTRO-PH 0211286;%%




\bibitem{ew300}
T.~Appelquist, H.~C.~Cheng and B.~A.~Dobrescu,
%``Bounds on universal extra dimensions,''
Phys.\ Rev.\ D {\bf 64}, 035002 (2001)
[arXiv:hep-ph/0012100];
%%CITATION = HEP-PH 0012100;%%
T.~Appelquist and H.~U.~Yee,
%``Universal extra dimensions and the Higgs boson mass,''
Phys.\ Rev.\ D {\bf 67}, 055002 (2003)
[arXiv:hep-ph/0211023].
%%CITATION = HEP-PH 0211023;%%

\bibitem{wmap}
C.~L.~Bennett {\it et al.},
%``First Year Wilkinson Microwave Anisotropy Probe (WMAP) 
Astrophys.\ J.\ Suppl.\  {\bf 148}, 1 (2003)
[arXiv:astro-ph/0302207].
%%CITATION = ASTRO-PH 0302207;%%


\bibitem{pythia}
T.~Sjostrand, P.~Eden, C.~Friberg, L.~Lonnblad, G.~Miu, S.~Mrenna and E.~Norrbin,
%``High-energy-physics event generation with PYTHIA 6.1,''
Comput.\ Phys.\ Commun.\  {\bf 135}, 238 (2001)
[arXiv:hep-ph/0010017].
%%CITATION = HEP-PH 0010017;%%

\bibitem{darksusy}
P.~Gondolo, J.~Edsj\"o, L.~Bergstr\"om, P.~Ullio, M.~Schelke and
E.~A.~Baltz, JCAP {\bf 07}, 008 (2004);
http://www.physto.se/~edsjo/darksusy/.




\bibitem{diffusion}
W.~R.~Webber, M.~A.~Lee and M.~Gupta,
Astrophys.\ J.{\bf 390} (1992) 96;
I.~V.~Moskalenko, A.~W.~Strong, S.~G.~Mashnik and J.~F.~Ormes,
 %``Challenging cosmic ray propagation with antiprotons: Evidence for a 'fresh'
%nuclei component?,''
Astrophys.\ J.\  {\bf 586}, 1050 (2003)
[arXiv:astro-ph/0210480];
%%CITATION = ASTRO-PH 0210480;%%
I.~V.~Moskalenko and A.~W.~Strong,
%``Positrons from particle dark-matter annihilation in the galactic halo:
%Propagation Green's functions,''
Phys.\ Rev.\ D {\bf 60}, 063003 (1999)
[arXiv:astro-ph/9905283].
%%CITATION = ASTRO-PH 9905283;%%







\bibitem{lossrate}
M.~S.~Longair,
{\it High Energy Astrophysics}, Cambridge University Press, New York, 1994.


\bibitem{nfw}
J.~F.~Navarro, C.~S.~Frenk and S.~D.~M.~White,
%``The Structure of Cold Dark Matter Halos,''
Astrophys.\ J.\  {\bf 462}, 563 (1996)
[arXiv:astro-ph/9508025].
%%CITATION = ASTRO-PH 9508025;%%

\bibitem{hooperpos}
D.~Hooper, J.~E.~Taylor and J.~Silk,
%``Can supersymmetry naturally explain the positron excess?,''
Phys.\ Rev.\ D {\bf 69}, 103509 (2004)
[arXiv:hep-ph/0312076].
%%CITATION = HEP-PH 0312076;%%

\bibitem{hess}
J.~A.~Hinton et al., New Astronomy Review {\bf 48}, 331 (2004).

\bibitem{veritas}
F.~Krennrich et al., New Astronomy Review {\bf 48}, 345 (2004).

\bibitem{magic}
E.~Lorenz et al., New Astronomy Review {\bf 48}, 339 (2004).

\bibitem{GLAST} N.~Gehrels \& P.~Michelson, Astropart.~Phys. {\bf 11}, 277
(1999).

\bibitem{amsreject}
G.~Carosi,
%``Positron / proton separation using the AMS-02 TRD,''
arXiv:physics/0409043.
%%CITATION = PHYS-ICS 0409043;%%


\end{thebibliography}
\end{document}